# III-V on CaF$_2$: a possible waveguiding platform for mid-IR photonic devices

**Ngoc-Linh Tran[1], Mario Malerba[1,*], Anne Talneau[1], Giorgio Biasiol[2], Oussama Ouznali[1], Adel Bousseksou[1], Jean-Michel Manceau[1], and Raffaele Colombelli[1]**

[1] *Centre de Nanosciences et de Nanotechnologies, CNRS UMR 9001, Université Paris-Sud, Université Paris-Saclay, C2N - Orsay, 91405 Orsay cedex, France*
[2] *Laboratorio TASC, CNR-IOM, Area Science Park, S.S. 14 km 163.5, Basovizza, I-34149 Trieste, Italy*
\* *mario.malerba@u-psud.fr*

**Abstract:** We developed a technique that enables to replace a metallic waveguide cladding with a low-index (n≈1.4) material – CaF$_2$ or BaF$_2$ – that in addition is transparent from the mid-IR up to the visible range: elevated confinement is preserved while introducing an optical entryway through the substrate. Replacing the metallic backplane also allows double-side patterning of the active region. Using this approach, we demonstrate strong light-matter coupling between an intersubband transition (λ~10 μm) and a dispersive resonator, at 300 K and at 78 K. Finally, we evaluate the potential of this approach as a platform for waveguiding in the mid-IR spectral range, with numerical simulations that reveal losses in the 1-10 cm$^{-1}$ range.

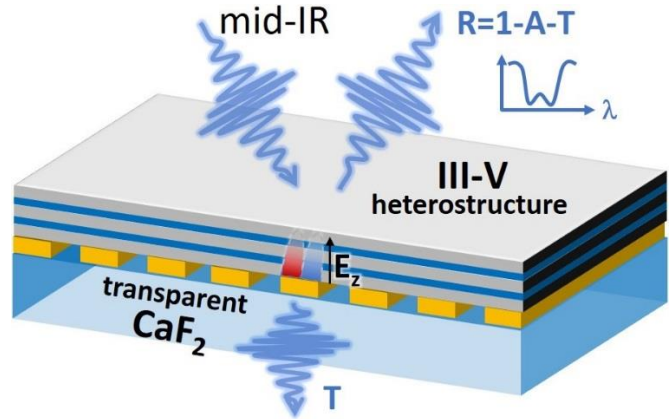

## 1. Introduction

The mid-IR is currently a spectral range of great technological interest. On one hand, in the field of optoelectronics, the control of material deposition at the atomic scale has revolutionized industrial applications, leading for instance to quantum cascade lasers (QCLs, from the mid-IR up to the THz spectral region) and quantum-well infrared photodetectors (QWIPs); on the other hand, by engineering plasmonic nanostructures, the ability to confine light down to the micro- and nano-scales has further boosted technological applications. To cite a few, enhanced optoelectronics [1] and energy-transfer [2], biochemical spectroscopy [3-6], electrochemical and photochemical catalysis [7], photovoltaics [8], and more. Novel materials have at last differentiated the possible platforms, allowing great manufacturing flexibility and possibilities [6, 9, 10].

In this context, and especially for actively controlled optoelectronic devices, such as detectors and emitters, confining and guiding light in high-index semiconductor layers is a basic, yet crucial functionality. When moderate confinements are required, slightly lower-index dielectric claddings are employed, in the near-IR and in the mid-IR spectral range as well [11], and very low losses are in general achieved. This is especially useful for passive devices for guidonics applications, for instance [12]. When instead elevated confinements are necessary, metallic claddings yield unsurpassed confinement factors, but higher losses must be factored in. However, they are low enough to be tolerated for longer mid-IR [13]and THz wavelengths [14]. This confinement strategy, that is typically based on Au-Au thermo-compression waferbonding, is amply used for active devices such THz lasers [15-17] and detectors [17-19], but also for mid-IR devices operating in the strong light-matter coupling regime [20-22].

Nevertheless, the choice of the cladding material does affect the device functionality well beyond the level of propagation losses introduced. For instance, metal-insulator-metal waveguides [23] – because of the metallic back-plane – totally prevent optical access of the device active core from the substrate, a functionality that would be extremely welcome for pump-probe experiments, especially in the emerging field of polaritonics [24]. Dielectric claddings present similar constraints, since they must be pseudomorphically grown on the active region and their energy gaps are typically larger than the active region one. Furthermore, the continuous metallic back-plane in traditional metal-insulator-metal architectures, acting as both bonding and cladding layer, constrains all patterning of photonic / electric nanostructures to take place on one side. This *de facto* blocks one of the two optical accesses to the active region (see Fig. 1(a)). It would be advantageous to employ as cladding material the one that best fits a specific application, independently of its chemical, epitaxial or thermal expansion compatibility with the active device core.

In this letter, we demonstrate an alternative approach to confine light inside high-refractive-index semiconductors as a possible platform for low-loss mid-IR waveguides. This approach is of interest for photonic devices in the mid-IR, a spectral range important for gas detection, spectroscopy, free space communications. It is based on the direct transfer of III-V material on $CaF_2$ (or equivalently $BaF_2$), and the permanent bonding to the new host substrate. We have chosen $CaF_2$ because it has an extremely low index of refraction in the mid-IR ($n \sim 1.3$), thus enabling elevated optical confinements, but also because it is transparent in the whole visible range.

On one hand, this latter peculiarity (the $CaF_2$ replaces the metallic back plane) enables pump-probe applications with pump wavelengths well into the visible/UV range, something that is of interest for the study of fundamental phenomena in intersubband polaritonics. In fact, a transparent substrate is a formidable asset to study the temporal construction of polaritonic states [24] and possibly observing dynamic Casimir radiation [25]. For this very reason we present here an experimental demonstration of the technology on grating-based mid-IR polaritonic devices [21].

On the other hand, the same architecture potentially enables other interesting families of devices. The optical access from both sides transforms it into a two-port system [26], with possible applications to devices exploiting coherent perfect absorption [27]. Also the possibility of patterning both sides of the active region, as well as further sequential lithography/bonding steps, may simplify the implementation of multi-layer metasurfaces (chiral structures for instance), or devices with microstructured front and back contacts (detectors, lasers). A last important feature is the possibility to reliably operate at cryogenic temperatures.

## 2. Methodological approach and fabrication

Today intense research activity is devoted to the development of low-loss waveguiding solutions in the infrared spectral range, mainly driven by the recent – yet extremely rapid – development of silicon photonics. In the 3-5 μm range, for instance, much effort has been put into realizing on-chip mid-IR biochemical sensors [28, 29], mostly exploiting silicon-on-insulator platforms [30, 31]. In parallel, germanium-on-insulator has been developed as an alternative, due to its broader transparency range, reaching the mid-IR up to 15 μm: Germanium-on-SiO2 [12, 32, 33] and Germanium-on-$CaF_2$ [34] for example. In most cases, materials are processed to achieve an intimate, direct adhesion between the active semiconductor region and the host substrate, mimicking the ideal case which would come from a direct epitaxial growth. This approach is interesting but limits applications to those materials that can be reliably wafer-bonded together, and hardly would allow a double-sided patterning of the thin active region – a possibility that can open interesting routes for electrically controlled active devices.

In the case we tackle in this work, III-V semiconductor on $CaF_2$ , traditional direct atomic-layer bonding is not straightforward because of the extremely different thermal expansion coefficients: $5.73 \times 10^{-6}$ °C$^{-1}$ for GaAs, against $18.85 \times 10^{-6}$ °C$^{-1}$ for $CaF_2$. Plasma-activated wafer bonding also appears difficult, due to the different nature of the Ca-$F_2$ (ionic) and III-V semiconductor (covalent) chemical bond. Alternative possibilities are: (i) use of an intermediate oxide layer; (ii) direct transfer exploiting Wan-der-Waals forces; and (iii) use of an intermediate bonding layer. Solution (i) is complicated by the lack of an oxide layer that can reliably bond to CaF$_2$ [35] and guarantee permanent adhesion, also when strong shear forces arise from cooling / heating the sample. Furthermore, the solution we look for must be flexible enough to be independent of the materials used, allowing us to bond also surfaces that have been previously patterned with optical nanostructures elements (i.e. for instance plasmonic elements or gratings). A non-flat patterned surface such as in this case can very hardly be pressure-bonded without affecting the quality and the geometry of the realized structures. We thus developed solutions (ii) and (iii), presented hereafter.

Optical grade CaF$_2$ substrates (thickness: 500 nm, surface quality S/D=40/20) were prepared following a standard cleaning procedure (sonication in warm acetone and rinsing in isopropanol, followed by 5' ashing in O$_2$ plasma).

The direct transfer procedure is described in Fig. 1(b). We proceeded with the following steps: (i) patterning of the coupler grating on the active region (standard electron-beam lithography, Ti/Au evaporation and liftoff – parameters in figure captions); (ii) spin coating of a protective PMMA A4 electron beam resist layer on the patterned active region (providing a clean interface at the end of the process) and fixing the sample to a temporary support substrate with a thermo-labile mounting adhesive (crystalbond 509); (iii) GaAs growth-substrate removal by mechanical polishing followed by chemical etch to AlGaAs stop layer; (iv) removal of stop layer in HF (49% conc.); (v) release of the Active Region (AR) in acetone, which dissolves PMMA and mounting adhesive. To obtain a clean and defect-less interface, minimizing destructive turbulence from weakly mixable solvents, we gradually substituted acetone with isopropyl alcohol and the latter with milli-Q water. The active region is at last allowed to rest on the substrate. After lifting it out of the liquid environment and allowing for evaporation, the patterned AR is inherently brought in close contact with the CaF$_2$ substrate by water's strong surface tension. Van der Waals interactions then keep it bonded to the new host substrate.

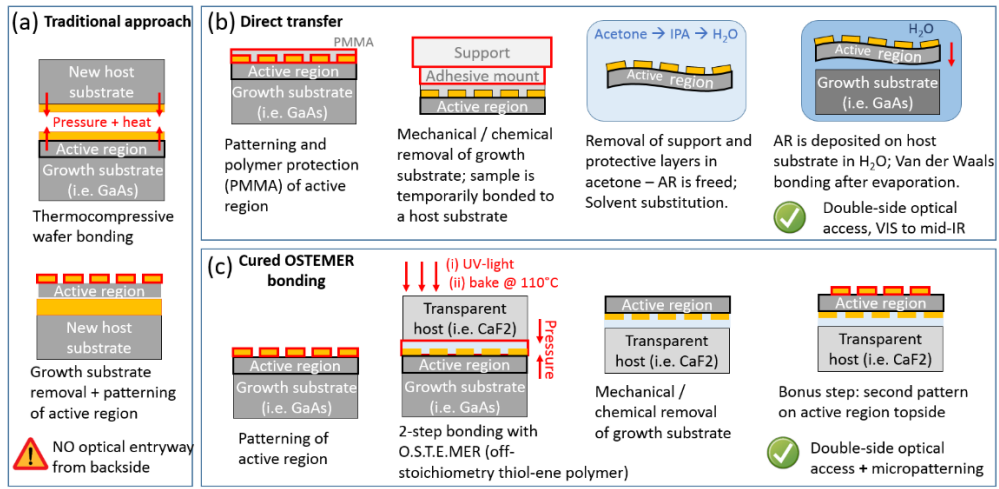

Fig. 1. Fabrication steps *(details available in the main text)*. (a) traditional approach involving thermocompressive waferbonding and substrate removal, yielding a double-metal optical cavity; (b) direct transfer of epitaxial layer via temporary encapsulation of the patterned top side, substrate removal, and deposition in liquid environment on the new transparent host substrate (bonding via Van der Waals interactions); (c) bonding with a commercial off-stoichiometry thiol-ene polymer (ostemer) and subsequent wafer removal, allowing micropatterning on both sides of the active region.

The advantage of this approach is a clean AR/CaF$_2$ interface. In contrast, further post-processing cannot be reliably performed, and the samples cannot be cooled down to cryogenic temperatures due to the weak interaction forces. We thus developed a more robust bonding strategy that is compatible with heating up to 180 °C (allowing further processing) and cooling down to liquid nitrogen temperature (-196 °C). The (acceptable) price to pay is the use of a thin polymeric bonding agent, whose robust thermo-mechanical properties however permit further cleanroom processing steps after transfer (optical or electron-beam lithography for instance), as well as device operation at cryogenic temperatures. Furthermore, its chemical formulation, based on SH- and OH- groups, makes it suitable for strong chemical bonding also with fluorides.

The intermediate bonding layer procedure is described in Fig. 1(c). The bonding polymer we chose is a commercially available off-stoichiometry thiol-ene agent used for microfluidics and micro-electromechanical systems (MEMS) (Ostemer 322 Crystal Clear, Mercene Labs, Stockholm, Sweden). To demonstrate post-processing compatibility, and double-side access to the active region, we tested two different layouts: grating patterning (via e-beam lithography and metallization) performed before (case 1), or after (case 2) bonding. In the former case, the grating is buried at the AR/CaF$_2$ interface; in the latter case, it lies at the AR/air top surface (after substrate removal). The detailed steps for case 1 are described in Fig. 1(c): (i) patterning of the coupler grating on the AR; (ii) 2-step bonding to CaF$_2$ (UV irradiation + soft bake hardening); (iii) substrate removal through mechanical polishing and chemical etch. Lithography and metal deposition, or any alternative processing step, can be performed at this point, in alternative or in addition to step (i).

## 3. Dispersion measurements in CaF$_2$-confined optical cavities

### 3.1. GaAs empty cavities and photonic confinement

Initially empty cavities have been implemented with both bonding procedures. The AR material consists of an undoped 1.5 μm-thick GaAs layer grown by Molecular Beam Epitaxy (MBE) onto a 500 nm-thick Al$_{0.5}$Ga$_{0.5}$As stop layer. The theoretical resonator dispersions obtained using a Rigorous Coupled Wave Analysis (RCWA) approach, along with the experimental validation, and the simulated electromagnetic modes confined within the optical cavities are reported in Fig. 2. The photonic band-structure of the devices are measured via angle-resolved reflectivity within a Fourier Transform interferometer (FTIR) as described in Ref.[21].

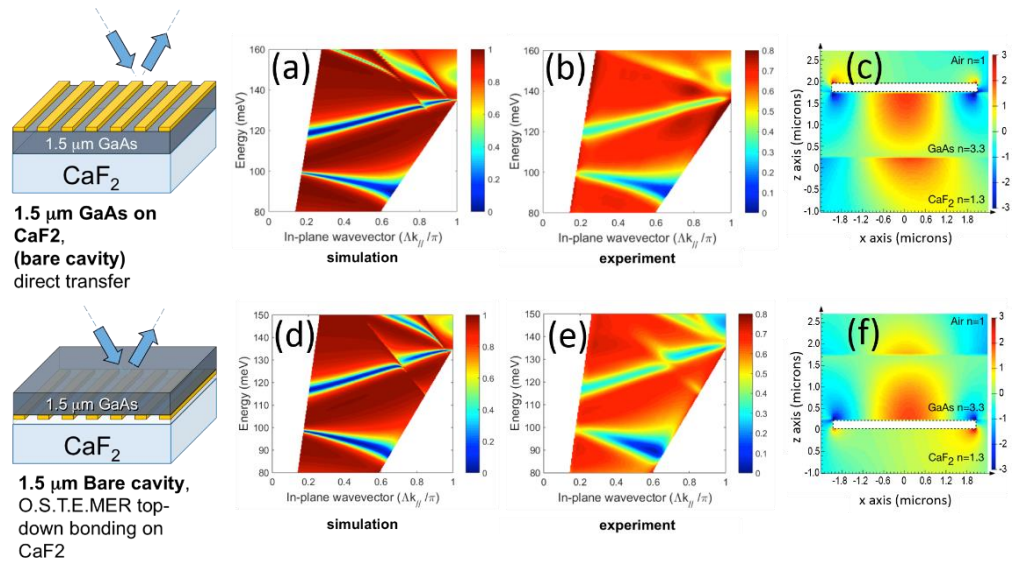

Fig. 2. Angle-resolved reflectivity experiments and simulations from bare cavities. Top panel: numerically calculated (a) and experimentally measured (b) dispersion of a 1.5 μm-thick undoped GaAs active region directly transferred on CaF$_2$ (grating up, on the air/GaAs interface); Bottom panel: simulated (d) and measured (e) dispersion of the same undoped GaAs cavity, ostemer-bonded to CaF$_2$ (grating down, buried on the GaAs / CaF$_2$ interface); FDTD calculation of E$_z$ field enhancement (z-axis perpendicular to active region) monitored in the region adjacent to the metal grating (white rectangle) for both grating-up (c) and grating-down (f) layouts.
*Grating parameters*: 4.8 μm period; 85% filling factor; 200 nm-thick Ti/Au metallization.

Figure 2 (upper panel, Fig. 2(a) and 2(b)) shows the classic band-structure of a fundamental TM00 mode folded in the first Brillouin zone. As in a traditional metal-metal layout, two photonic branches are separated by a photonic band-gap at the 2$^{nd}$ order Bragg condition. Differently from a metal-metal structure, however, an additional light line crosses the upper photonic branch. It originates from the low-index CaF$_2$ substrate. Similar dispersions (lower panel, Figs. 2(d) and 2(e)) are measured, and numerically confirmed, from the more unusual grating-down layout, where the AR is tightly bonded to the host CaF$_2$ substrate with the bonding polymer. FDTD simulations of the E$_z$ field component (Figs. 2(c) and 2(f)) performed on both layouts show symmetric EM modes and field confinement. Note: positioning the coupler grating at the GaAs/CaF$_2$ interface improves the confinement, as the difference in refractive index with the other free interface ($n_{air} = 1$ vs. $n_{GaAs} = 3.3$) is the highest achievable. The total damping rates of the photonic modes are similar for the two architectures and also similar to the ones obtained with a metal-metal approach, with total Q-factors ($Q_{tot}$) of the order of 20 in both cases. When the active core is undoped, $Q_{tot}$ in a metal-AR-grating system is dominated by ohmic losses in the metallic layers, while in the current III-V-on-CaF$_2$ architecture radiative losses play an important role. However, when the active region is doped (i.e. there are losses) it can be shown that most of absorption takes place in the AR in both cases (see further in text).

### *3.2. GaAs quantum-wells and strong coupling*

We have then explored if the strong light-matter coupling regime can be attained by replacing the bare GaAs active region with a repetition of MBE-grown doped semiconductor quantum wells (QW) of similar total thickness. The heterostructure consists of 50 repetitions of 8.3 nm-wide GaAs QWs separated by 20-nm-thick Al$_{0.33}$Ga$_{0.67}$As barriers. The latter ones are nominally δ-doped to n$_{2D}$= 1.34x10$^{12}$ cm$^{-2}$ at the center. The AR exhibits a resonant intersubband (ISB) transition absorption at 118 meV at 300 K, as measured with multi-pass transmission experiments [21]. Figure 3 presents the polaritonic dispersions for the two different grating layouts. Both are obtained with the polymer bonding approach, but one (panel a) features a buried grating, while the second one (panel b) a surface one. Details of the gratings are in the caption. With the ISB transition active thanks to the doping, new eigenmodes appear - separated by twice the Rabi energy at the anti-crossing – that are commonly named intersubband polaritons. They are the classic signature of a system operating in strong coupling.

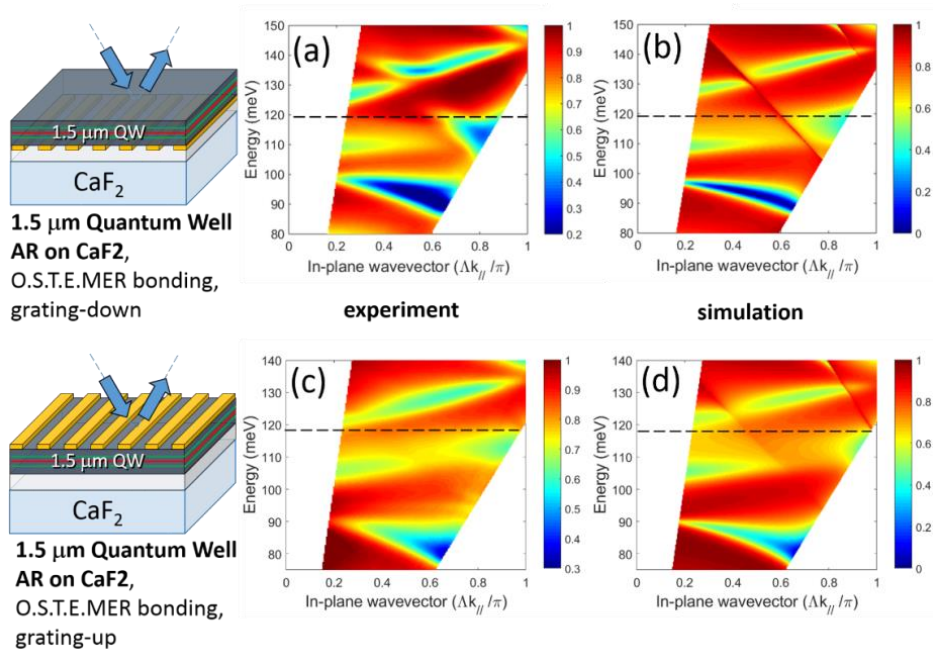

Fig. 3. Angle-resolved reflectivity experiments and simulations from quantum-well active regions, comparing the polaritonic mixed mode on (a) grating-down layout and (b) grating-up standard layout. Both specimens are bonded with ostemer. The photonic cavity mode is strongly coupled to the quantum well ISB transition at 119 meV, splitting into lower and upper polariton branches. The possibility of performing lithographic steps both before and after bonding outlooks double side metal-AR-metal nanopatterning.

*Grating parameters*: (top panel) 4.95 μm period, 83.3% filling factor, (bottom panel) 5.35 μm period, 84% filling factor. 200 nm-thick Ti/Au metallization.

The polaritonic splitting is slightly larger in the buried grating sample: this is correctly reproduced in the numerical simulations (Figs. 3(b) and 3(d)). The origin must be sought in the electromagnetic overlap factor, which is larger in the buried grating sample than in the top grating one (see FDTD calculations in Fig. 2), as the field can only leak trough the grating openings, while being confined above from air. It confirms therefore that burying the grating at the III-V/$CaF_2$ interface is advantageous. It also permits to more uniformly excite the AR with a near-IR laser beam for perspective pump-probe experiments [24].

Numerical simulations are in very good agreement with the experiments (optical parameters *n, k* of ostemer are obtained from the manufacturer or deduced from absorption measurements, as detailed in the following paragraph). Furthermore, the value of the CaF2 permittivity can be fine-tuned thanks to the position of the "$CaF_2$ light line" that crosses the polaritonic branches.

## 4. Cryogenic performance, and device-oriented perspectives

A crucial feature is the robustness of the bonded samples against operation at cryogenic temperatures. The choice of the bonding agent impacts considerably on the final use that can be envisioned for the presented platform. Brittle or glassy bonding agents (i.e. most resist polymers) will not withstand cooling to cryogenic temperatures, eventually tearing the active region. Weakly interacting polymers, such as elastomers (i.e. for instance PDMS), are mechanically more deformable, but they suffer from weak chemical interaction, eventually leading to AR detachment after few heating / cooling cycles. The material we used in this paper presents both characteristics (elastomeric mechanical response and strong covalent bonds). To demonstrate the compatibility for devices operating at liquid nitrogen temperatures, i.e. quantum well infrared detectors for instance, we mounted the samples on copper blocks, placed them into an open flow cryostat and cooled down to 78 K. The cooling/warm-up cycle was repeated a few times: no sample degradation was observed. Figure 4(a) shows the reflectivity of the sample at normal incidence ($k_{//}=0$) for both 300 K and 78 K.

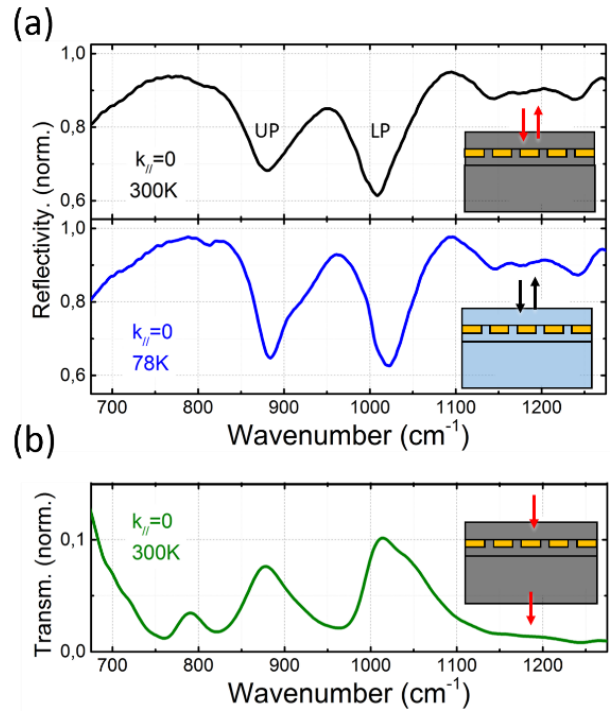

Fig. 4. Reflectivity experiments in the grating-down configuration (sample depicted in Fig. 3(a)) with normal incidence, TM polarization, at room (above) and liquid nitrogen (below) temperature; (b) Transmission experiment for the same sample, showing an additional purely photonic peak (hidden in reflection) at 790 cm$^{-1}$ (98 meV, compare to Fig. 3(a)), besides the expected lower and upper polariton excitation. Absorption is calculated as: A = 1- R-T = 0.24 (UP), 0.28 (UP).

Cooling induces a slight linewidth narrowing and spectral blue-shift of the polaritonic resonances. At last, since one of the desired features of III-V-on-$CaF_2$ bonding is substrate mid-IR transparency, we probed the transmittance (2nd port of the system) at normal incidence in order to assess the level of absorption of the system. Figure 4(b) reveals the two polaritonic states in transmission at 300 K, whose spectral position is in excellent agreement with the reflectivity measurements. Most importantly, it is possible to estimate the system absorption as A = 1 – R – T: 24% and 28% of the light is absorbed respectively by the upper and lower polaritonic resonances. This value is in line with what is achieved with systems with a metallic back-plane when not in critical coupling [36], and it can be improved with judicious sample design.

Having experimentally established that this III-V on $CaF_2$ bonding approach suits applications for polaritonic devices, also at cryogenic temperatures, we analyzed with numerical simulations its suitability in the field of guidonics in the mid-IR spectral range. This is of interest for passive and active devices [12, 37, 38], but also in principle for mid-IR frequency conversion applications [39-41].

We have numerically simulated the propagation of a $TM_{00}$ mode into an etched GaAs passive waveguide bonded onto a $CaF_2$ substrate, as schematized in Fig. 5(a).

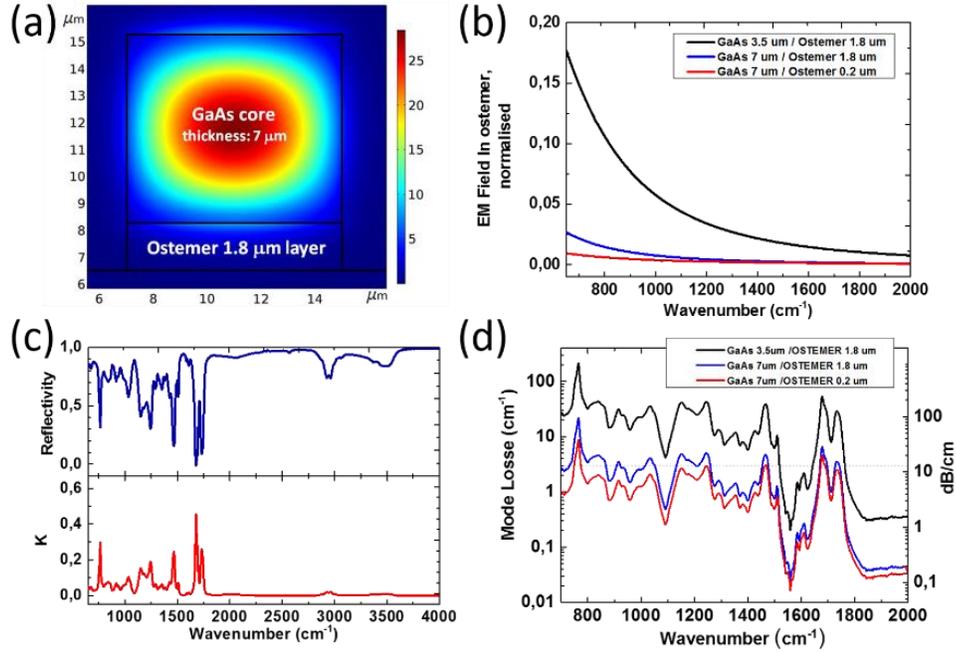

Fig. 5. – (a) Field distribution ($|E_z|^2$ is plotted) of the $TM_{00}$ mode in a semi-insulating GaAs-on-$CaF_2$ waveguide; (b) electromagnetic overlap with the ostemer layer of the $TM_{00}$ mode as a function of wavenumber, for different GaAs and ostemer thicknesses. The plotted quantity is $\iint_{Ostemer} |\boldsymbol{E}|^2 / \iint_{Full\ domain} |\boldsymbol{E}|^2$;

(c) reflectivity spectrum from a 1.8 μm-thick ostemer layer deposited on a gold-coated substrate and deduced imaginary part of the refractive index; (d) propagation losses for mode TM00 as a function of the wavelength, for different GaAs and ostemer thicknesses. The ridge width is 8 μm. Above 1800 cm$^{-1}$ losses are below 2 dB/cm (0.46 cm$^{-1}$). In the region of ostemer ro-vibrational transitions, it is possible to keep the losses below in the 1-3 cm$^{-1}$ with a reasonably thick AR.

The modal overlap factor with the AR is always above 90%, thanks to the low index of refraction of both the bonding polymer and $CaF_2$. The propagation losses are set by the field penetration in the ostemer layer, that is plotted in Fig. 5(b) – as a function of the wavenumber – for two different AR and ostemer thicknesses. The calculation reveals that the AR thickness plays a primary role, while decreasing the ostemer thickness has an almost negligible effect above 1000 cm$^{-1}$.

Fig. 5(c) (top panel) shows the experimental reflectivity of a 1.8 μm-thick layer of ostemer, spin casted on a gold-coated glass slide. Fig. 5(c) (bottom panel) shows the inferred imaginary part of ostemer refractive index, $k$, deduced from the reflectivity measurement. The absorption is first evaluated as (1-R) and is then normalized to the sample thickness, in order to obtain the absorption coefficient per unit length $\alpha$. Finally, the imaginary part of the refractive index $k$ is inferred as $k = \lambda \alpha / 4\pi$ and fed into the numerical simulations' material properties. The real part $n$ of the refractive index was taken to be around 1.6, as declared by the manufacturer for the visible range.

Fig. 5(d) reports the propagation losses of the $TM_{00}$ mode as a function of the wavenumber for the three different configurations. The dotted line corresponds to losses of 3 cm$^{-1}$. For a 7 μm-thick waveguide the peak losses lie in the 3-5 cm$^{-1}$ range above 1000 cm$^{-1}$ wavenumber. If the ostemer thickness is further reduced to 0.2 μm, the situation improves with peak losses in the 1-3 cm$^{-1}$ range. Note also that deep oscillations are present and losses can reach values below 0.5 cm$^{-1}$ for specific frequency ranges: above 1800 cm$^{-1}$ for instance losses become negligible.

These results show that this platform is compatible with the implementation of passive waveguides in mid-IR-on-a-chip applications or frequency conversion above 1800 cm$^{-1}$. Below this value, further improvements would be needed to reach propagation losses below 1 dB/cm. On the other hand, the estimated level of loss is compatible with the monolithic integration of mid-IR active devices [38] as high AR overlap factors (> 90%) can be obtained.

## 5. Conclusions

We have developed a technique that permits to replace the metallic cladding with a low-index material ($CaF_2$ or $BaF_2$), which is transparent from the mid-IR up to the visible range: elevated confinement is preserved, while introducing an optical entryway through the substrate along with active region double side micro-structuring potential. With this approach we have demonstrated GaAs-based optical cavities transferred on $CaF_2$ both directly or with an intermediate bonding polymer. The use of judiciously doped semiconductor QWs led to the observation of strong light-matter coupling between an ISB transition and the resonator photonic mode, at 300 K and at 78 K. We have also shown that placing the coupler grating on the air/active region boundary or burying it on the active region/$CaF_2$ boundary leads to equivalent results. In the field of polaritonics such a two port system [27, 36], with a near-IR transparent substrate and flexibility in radiation coupling, can represent a precious asset to study the temporal construction of polaritonic states [24] and possibly Casimir radiation. Especially for the latter application, the active region illumination through the $CaF_2$ offers a homogeneous excitation of the QWs, hence by-passing problems of charge diffusion that occurs when illuminating from the grating side.

**Funding**
We acknowledge financial support from the European Research Council (IDEAS-ERC) ("GEM") (306661) and from the European Union's Horizon 2020 research and innovation program under the Marie Sklodowska-Curie grant agreement No 748071.

**Acknowledgments**
We thank K. Pantzas and I. Sagnes for useful discussions, and E. Hertz for his help in cleanroom processes. This work was partly supported by the French RENATECH network.

**References**

1. M. Malerba, T. Ongarello, B. Paulillo, J.-M. Manceau, G. Beaudoin, I. Sagnes, F. D. Angelis, and R. Colombelli, "Towards strong light-matter coupling at the single-resonator level with sub-wavelength mid-infrared nano-antennas," Applied Physics Letters **109**, 021111 (2016).
2. D. Garoli, E. Calandrini, A. Bozzola, M. Ortolani, S. Cattarin, S. Barison, A. Toma, and F. De Angelis, "Boosting infrared energy transfer in 3D nanoporous gold antennas," Nanoscale **9**, 915-922 (2017).
3. J. Haas and B. Mizaikoff, "Advances in Mid-Infrared Spectroscopy for Chemical Analysis," Annual Review of Analytical Chemistry **9**, 45-68 (2016).
4. E. Shkondin, T. Repän, M. E. Aryaee Panah, A. V. Lavrinenko, and O. Takayama, "High Aspect Ratio Plasmonic Nanotrench Structures with Large Active Surface Area for Label-Free Mid-Infrared Molecular Absorption Sensing," ACS Applied Nano Materials **1**, 1212-1218 (2018).
5. R. La Rocca, G. C. Messina, M. Dipalo, V. Shalabaeva, and F. De Angelis, "Out-of-Plane Plasmonic Antennas for Raman Analysis in Living Cells," Small **11**, 4632-4637 (2015).
6. G. C. Messina, M. Malerba, P. Zilio, E. Miele, M. Dipalo, L. Ferrara, and F. De Angelis, "Hollow plasmonic antennas for broadband SERS spectroscopy," Beilstein Journal of Nanotechnology **6**, 492-498 (2015).
7. Y. Sang, H. Liu, and A. Umar, "Photocatalysis from UV/Vis to Near-Infrared Light: Towards Full Solar-Light Spectrum Activity," ChemCatChem **7**, 559-573 (2015).

8. A. I. Yakimov, V. V. Kirienko, A. A. Bloshkin, V. A. Armbrister, A. V. Dvurechenskii, and J. M. Hartmann, "Photovoltaic Ge/SiGe quantum dot mid-infrared photodetector enhanced by surface plasmons," Opt. Express **25**, 25602-25611 (2017).
9. Y. Zhong, S. D. Malagari, T. Hamilton, and D. M. Wasserman, "Review of mid-infrared plasmonic materials," Journal of Nanophotonics **9**, 21 (2015).
10. S. Law, V. Podolskiy, and D. Wasserman, "Towards nano-scale photonics with micro-scale photons: the opportunities and challenges of mid-infrared plasmonics," Nanophotonics **2**, 103 (2013).
11. G. Claire, C. Federico, L. S. Deborah, and Y. C. Alfred, "Recent progress in quantum cascade lasers and applications," Reports on Progress in Physics **64**, 1533 (2001).
12. J. M. Ramirez, Q. Liu, V. Vakarin, J. Frigerio, A. Ballabio, X. Le Roux, D. Bouville, L. Vivien, G. Isella, and D. Marris-Morini, "Graded SiGe waveguides with broadband low-loss propagation in the mid infrared," Opt. Express **26**, 870-877 (2018).
13. D. Chastanet, A. Bousseksou, G. Lollia, M. Bahriz, F. H. Julien, A. N. Baranov, R. Teissier, and R. Colombelli, "High temperature, single mode, long infrared ($\lambda$ = 17.8 μm) InAs-based quantum cascade lasers," Applied Physics Letters **105**, 111118 (2014).
14. Y. Chassagneux, R. Colombelli, W. Maineult, S. Barbieri, S. P. Khanna, E. H. Linfield, and A. G. Davies, "Graded photonic crystal terahertz quantum cascade lasers," Applied Physics Letters **96**, 031104 (2010).
15. B. S. Williams, "Terahertz quantum-cascade lasers," Nature Photonics **1**, 517 (2007).
16. G. Xu, L. Li, N. Isac, Y. Halioua, A. G. Davies, E. H. Linfield, and R. Colombelli, "Surface-emitting terahertz quantum cascade lasers with continuous-wave power in the tens of milliwatt range," Applied Physics Letters **104**, 091112 (2014).
17. D. Bachmann, M. Rösch, M. J. Süess, M. Beck, K. Unterrainer, J. Darmo, J. Faist, and G. Scalari, "Short pulse generation and mode control of broadband terahertz quantum cascade lasers," Optica **3**, 1087-1094 (2016).
18. B. Paulillo, S. Pirotta, H. Nong, P. Crozat, S. Guilet, G. Xu, S. Dhillon, L. H. Li, A. G. Davies, E. H. Linfield, and R. Colombelli, "Ultrafast terahertz detectors based on three-dimensional meta-atoms," Optica **4**, 1451-1456 (2017).
19. Y. N. Chen, Y. Todorov, B. Askenazi, A. Vasanelli, G. Biasiol, R. Colombelli, and C. Sirtori, "Antenna-coupled microcavities for enhanced infrared photo-detection," Applied Physics Letters **104**, 031113 (2014).
20. A. Delteil, A. Vasanelli, Y. Todorov, C. Feuillet Palma, M. Renaudat St-Jean, G. Beaudoin, I. Sagnes, and C. Sirtori, "Charge-Induced Coherence between Intersubband Plasmons in a Quantum Structure," Physical Review Letters **109**, 246808 (2012).
21. M. Helm, "The basic physics of intersubband transitions," in *Semiconductors and semimetals* (Elsevier, 1999), pp. 1-99.
22. J.-M. Manceau, N.-L. Tran, G. Biasiol, T. Laurent, I. Sagnes, G. Beaudoin, S. D. Liberato, I. Carusotto, and R. Colombelli, "Resonant intersubband polariton-LO phonon scattering in an optically pumped polaritonic device," Applied Physics Letters **112**, 191106 (2018).
23. B. S. Williams, S. Kumar, H. Callebaut, Q. Hu, and J. L. Reno, "Terahertz quantum-cascade laser at $\lambda \approx$100 μm using metal waveguide for mode confinement," Applied Physics Letters **83**, 2124-2126 (2003).
24. G. Günter, A. A. Anappara, J. Hees, A. Sell, G. Biasiol, L. Sorba, S. De Liberato, C. Ciuti, A. Tredicucci, A. Leitenstorfer, and R. Huber, "Sub-cycle switch-on of ultrastrong light–matter interaction," Nature **458**, 178 (2009).

25. C. Ciuti, G. Bastard, and I. Carusotto, "Quantum vacuum properties of the intersubband cavity polariton field," Physical Review B **72**, 115303 (2005).
26. H. A. Haus, *Waves and fields in optoelectronics* (Prentice-Hall, 1984).
27. S. Zanotto, F. P. Mezzapesa, F. Bianco, G. Biasiol, L. Baldacci, M. S. Vitiello, L. Sorba, R. Colombelli, and A. Tredicucci, "Perfect energy-feeding into strongly coupled systems and interferometric control of polariton absorption," Nature Physics **10**, 830 (2014).
28. M. Sieger and B. Mizaikoff, "Toward On-Chip Mid-Infrared Sensors," Analytical Chemistry **88**, 5562-5573 (2016).
29. D. A. Mohr, D. Yoo, C. Chen, M. Li, and S.-H. Oh, "Waveguide-integrated mid-infrared plasmonics with high-efficiency coupling for ultracompact surface-enhanced infrared absorption spectroscopy," Opt. Express **26**, 23540-23549 (2018).
30. Y. Zou, S. Chakravarty, C.-J. Chung, X. Xu, and R. T. Chen, "Mid-infrared silicon photonic waveguides and devices [Invited]," Photon. Res. **6**, 254-276 (2018).
31. Y. Chen, H. Lin, J. Hu, and M. Li, "Heterogeneously Integrated Silicon Photonics for the Mid-Infrared and Spectroscopic Sensing," ACS Nano **8**, 6955-6961 (2014).
32. J. Kang, M. Takenaka, and S. Takagi, "Novel Ge waveguide platform on Ge-on-insulator wafer for mid-infrared photonic integrated circuits," Opt. Express **24**, 11855-11864 (2016).
33. D. Marris-Morini, V. Vakarin, M. Ramirez Joan, Q. Liu, A. Ballabio, J. Frigerio, M. Montesinos, C. Alonso-Ramos, X. Le Roux, S. Serna, D. Benedikovic, D. Chrastina, L. Vivien, and G. Isella, "Germanium-based integrated photonics from near- to mid-infrared applications," in *Nanophotonics,* (2018), p. 1781.
34. S. Kim, J.-H. Han, J.-P. Shim, H.-j. Kim, and W. J. Choi, "Verification of Ge-on-insulator structure for a mid-infrared photonics platform," Opt. Mater. Express **8**, 440-451 (2018).
35. Y. Wu, J. T. Mayer, E. Garfunkel, and T. E. Madey, "X-ray Photoelectron Spectroscopy Study of Water Adsorption on BaF2(111) and CaF2(111) Surfaces," Langmuir **10**, 1482-1487 (1994).
36. J.-M. Manceau, S. Zanotto, I. Sagnes, G. Beaudoin, and R. Colombelli, "Optical critical coupling into highly confining metal-insulator-metal resonators," Applied Physics Letters **103**, 091110 (2013).
37. A. Spott, J. Peters, M. L. Davenport, E. J. Stanton, C. D. Merritt, W. W. Bewley, I. Vurgaftman, C. S. Kim, J. R. Meyer, J. Kirch, L. J. Mawst, D. Botez, and J. E. Bowers, "Quantum cascade laser on silicon," Optica **3**, 545-551 (2016).
38. S. Jung, J. Kirch, J. H. Kim, L. J. Mawst, D. Botez, and M. A. Belkin, "Quantum cascade lasers transfer-printed on silicon-on-sapphire," Applied Physics Letters **111**, 211102 (2017).
39. A. Grisard, E. Lallier, and B. Gérard, "Quasi-phase-matched gallium arsenide for versatile mid-infrared frequency conversion," Opt. Mater. Express **2**, 1020-1025 (2012).
40. Q. Clément, J. M. Melkonian, J. B. Dherbecourt, M. Raybaut, A. Grisard, E. Lallier, B. Gérard, B. Faure, G. Souhaité, and A. Godard, "Longwave infrared, single-frequency, tunable, pulsed optical parametric oscillator based on orientation-patterned GaAs for gas sensing," Opt. Lett. **40**, 2676-2679 (2015).
41. K. L. Vodopyanov, O. Levi, P. S. Kuo, T. J. Pinguet, J. S. Harris, M. M. Fejer, B. Gerard, L. Becouarn, and E. Lallier, "Optical parametric oscillation in quasi-phase-matched GaAs," Opt. Lett. **29**, 1912-1914 (2004).